\documentclass[10pt,twocolumn,aps,pra,showpacs]{revtex4}

\usepackage{bbm}

\begin{document}

\title{On the class of diffusion operators for fast quantum search}         
\author{Avatar Tulsi\\
        {\small Department of Physics, IIT Bombay, Mumbai-400076, India}}  

\email{tulsi9@gmail.com} 

\begin{abstract}

Grover's quantum search algorithm evolves a quantum system from a known source state $|s\rangle$ to an unknown target state $|t\rangle$ using the selective phase inversions, $I_{s}$ and $I_{t}$, of these two states. In one of the generalizations of Grover's algorithm, $I_{s}$ is replaced by a general diffusion operator $D_{s}$ having $|s\rangle$ as an eigenstate and $I_{t}$ is replaced by a general selective phase rotation $I_{t}^{\phi}$. A fast quantum search is possible as long as the operator $D_{s}$ and the angle $\phi$ satisfies certain conditions. These conditions are very restrictive in nature. Specifically, suppose $|\ell\rangle$ denote the eigenstates of $D_{s}$ corresponding to the eigenphases $\theta_{\ell}$. Then the sum of the terms $|\langle \ell|t\rangle|^{2}\cot(\theta_{\ell}/2)$ over all $\ell \neq s$ has to be almost equal to $\cot(\phi/2)$ for a fast quantum search. In this paper, we show that this condition can be significantly relaxed by introducing appropriate modifications of the algorithm. This allows access to a more general class of diffusion operators for fast quantum search. 

\end{abstract}


\maketitle

\section{INTRODUCTION}

Grover's quantum search algorithm or more generally quantum amplitude amplification evolves a quantum system from a known \emph{source} state $|s\rangle$ to an unknown but desired \emph{target} state $|t\rangle$~\cite{grover,qaa1,qaa2}. It does so by using selective phase inversion operators, $I_{s}$ and $I_{t}$, of these two quantum states. The algorithm iteratively applies the operator $\mathcal{A}(s,t) = -I_{s}I_{t}$ on $|s\rangle$ to get $|t\rangle$. The required number of iterations is $O(1/\alpha)$ where $\alpha = |\langle t|s\rangle|$. For search problem, $|s\rangle$ is chosen to be the uniform superposition of all $N$ basis states to be searched i.e. $|s\rangle = \sum_{i}|i\rangle/\sqrt{N}$. In case of a unique solution, the target state $|t\rangle$ is a unique basis state and $\alpha = |\langle t|s\rangle| = 1/\sqrt{N}$. Thus Grover's algorithm outputs a solution in just $O(\sqrt{N})$ time steps whereas \emph{classical} search algorithms take $O(N)$ time steps to do so. Quantum search algorithm and amplitude amplification are proved to be strictly optimal~\cite{optimal}.

A generalization of quantum search algorithm was presented in ~\cite{general}. The general quantum search algorithm (hereafter referred to as general algorithm) replaces $I_{s}$ by a more general diffusion operator $D_{s}$ with the only restriction of having $|s\rangle$ as an eigenstate. This restriction is reasonable as the diffusion operator should have some special connection with the source state. Let the normalized eigenspectrum of $D_{s}$ be given by $D_{s}|\ell\rangle = e^{\imath\theta_{\ell}}|\ell\rangle$ with $|\ell\rangle$ as the eigenstates and $e^{\imath\theta_{\ell}}$ ($\theta_{\ell}$) as the corresponding 
eigenvalues (eigenphases). We choose $D_{s}|s\rangle = |s\rangle$, i.e. $\theta_{\ell=s} = 0$. The general algorithm also replaces $I_{t}$ by a general selective phase rotation $I_{t}^{\phi}$ which multiplies $|t\rangle$ by a phase factor of $e^{\imath \phi}$ but leaves all states orthogonal to $|t\rangle$ unchanged. When $\phi$ is $\pi$, $I_{t}^{\phi}$ becomes $I_{t}$. Thus the general algorithm iterates the operator $\mathcal{S} = D_{s}I_{t}^{\phi}$ on $|s\rangle$ and its dynamics can be understood by analyzing the eigenspectrum of $\mathcal{S}$. 

This analysis was done in ~\cite{general} and we found that the performance of the general algorithm depends upon the moments $\Lambda_{1}$ and $\Lambda_{2}$ given by
\begin{equation}
\Lambda_{p} = \sum_{\ell \neq s}|\langle \ell|t\rangle|^{2}\cot^{p}\frac{\theta_{\ell}}{2}. \label{momentdefine}
\end{equation}
Thus $\Lambda_{p}$ is the $p^{\rm th}$ moment of $\cot\frac{\theta_{\ell}}{2}$ with respect to the distribution $|\langle \ell |t\rangle|^{2}$ over all $\ell \neq s$. Using these moments, we can define two quantities $A$ and $B$ as 
\begin{equation}
A = \Lambda_{1} - \cot \frac{\phi}{2},\ \ B = \sqrt{1+\Lambda_{2}}. \label{ABdefine}
\end{equation} 
It has been shown in ~\cite{general} that a fast general algorithm is possible if and only if $A = O(\alpha B) \approx 0$ as typically $\alpha \ll 1$. Thus $\sum_{\ell \neq s}|\langle \ell|t\rangle|^{2}\cot\frac{\theta_{\ell}}{2}$ must be almost equal to $\cot \frac{\phi}{2}$ for a fast quantum search. This is a very restrictive condition. In this paper, we present a modification of the general algorithm which does not require this kind of restrictive condition for its success. With this modification, the general algorithm becomes significantly flexible and works with a more general class of diffusion operators. Thus it allows for a successful quantum search in more general situations.

In next section, we present a brief review of the general algorithm. In Section III, we present the modification of the general algorithm. In Section IV, we discuss possible applications and conclude the paper.

\section{GENERAL ALGORITHM}

We briefly review the general algorithm~\cite{general}. It iterates the operator $\mathcal{S} = D_{s}I_{t}^{\phi}$ on $|s\rangle$. For simplicity, $|s\rangle$ is assumed to be a non-degenerate eigenstate of $D_{s}$ with eigenvalue $1$. The normalized eigenspectrum of $D_{s}$ is given by $D_{s}|\ell\rangle = e^{\imath\theta_{\ell}}|\ell\rangle$. We have $\theta_{\ell = s} = 0$. Let other 
eigenvalues satisfy
\begin{equation}
|\theta_{\ell \neq s}| \geq \theta_{\rm min} > 0,\ \ \theta_{\ell} \in [-\pi,\pi]. \label{othereigenvalues}
\end{equation} 
We need to find the eigenspectrum of $\mathcal{S}$ to analyse its iteration on $|s\rangle$. The secular equation was found in ~\cite{general} to be 
\begin{equation}
\sum_{\ell}|\langle \ell|t\rangle|^{2}\cot\frac{\lambda_{k}-\theta_{\ell}}{2} = \cot\frac{\phi}{2}. \label{secular}
\end{equation}
Any eigenvalue $e^{\imath \lambda_{k}}$ of $\mathcal{S}$ has to satisfy above equation.

Since $\cot x$ varies monotonically with $x$ except for the jump from $-\infty$ to $\infty$ when $x$ crosses zero, there is a unique solution $\lambda_{k}$ between each pair of consecutive $\theta_{\ell}$'s. As $\theta_{\ell = s} = 0$, there can be at most two solutions $\lambda_{k}$ in the interval $[-\theta_{\rm min},\theta_{\rm min}]$. Let these two solutions be $\lambda_{\pm}$. We have $|\lambda_{\pm}| < \theta_{\rm min}$. The two eigenstates $|\lambda_{\pm}\rangle$ corresponding to these two eigenvalues $e^{\imath \lambda_{\pm}}$ are the only relevant eigenstates for our algorithm as the initial state $|s\rangle$ is almost completely spanned by these two eigenstates provided we assume $|\lambda_{\pm}| \ll \theta_{\rm min}$. 

As shown in ~\cite{general}, the eigenphases $\lambda_{\pm}$ are given by 
\begin{equation}
\lambda_{\pm} = \pm\frac{2\alpha}{B}(\tan \eta)^{\pm 1}\  ;\  \cot 2\eta = \frac{A}{2\alpha B}\ .\label{solutions}
\end{equation}
where $\eta$ is chosen to be within the interval $[0,\pi/2]$ and the quantities $A$ and $B$ are as defined in Eq. (\ref{ABdefine}).

Eq. (20) of ~\cite{general} gives us the target state $|t\rangle$ in terms of two relevant eigenstates $|\lambda_{\pm}\rangle$. We have
\begin{equation}
|t\rangle = \frac{|w\rangle}{B|\sin \frac{\phi}{2}|} + |\lambda_{\perp}\rangle,\ \ |w\rangle = \sin \eta |\lambda_{+}\rangle + \cos \eta |\lambda_{-}\rangle,
\end{equation}
where $|w\rangle$ is the normalized projection of $|t\rangle$ on the $|\lambda_{\pm}\rangle$-subspace, and $|\lambda_{\perp}\rangle$ is a state orthogonal to this subspace.

Eq. (23) and (24) of ~\cite{general} gives us the initial state $|s\rangle$ and the effect of iterating $\mathcal{S}$ on $|s\rangle$ in terms of two relevant eigenstates $|\lambda_{\pm}\rangle$. We have
\begin{equation}
|s\rangle = e^{-\imath \phi/2}[e^{\imath \lambda_{+}/2}\cos \eta|\lambda_{+}\rangle - e^{\imath \lambda_{-}/2} \sin \eta|\lambda_{-}\rangle], \label{slambdapmexpansion}
\end{equation}
and
\begin{equation}
\mathcal{S}^{q}|s\rangle =  e^{-\imath \phi/2} [e^{\imath q'\lambda_{+}} \cos \eta|\lambda_{+}\rangle - e^{\imath q'\lambda_{-}}\sin \eta|\lambda_{-}\rangle],
 \label{stateexpand}
\end{equation}
where $q'  = q+\frac{1}{2}$.

The success probability of the algorithm is the probability of obtaining $|t\rangle$ upon measuring $\mathcal{S}^{q}|s\rangle$ which is $|\langle t|\mathcal{S}^{q}|s\rangle|^{2}$. Let this probability obtain its first maximum for $q=q_{\rm m}$. Let us define a state $|u\rangle$ such that $|u\rangle = \mathcal{S}^{q_{\rm m}}|s\rangle$. Then, by definition, the maximum success probability is 
\begin{equation}
P_{\rm m} = \beta^{2},\ \ \beta = |\langle t|u\rangle|.
\end{equation}
Eq. (27) of ~\cite{general} gives $q_{\rm m}$ and $\beta$ as  
\begin{equation}
q_{\rm m} \approx \frac{\pi B\sin 2\eta}{4\alpha},\ \ \beta = \frac{\sin 2\eta}{B \sin \frac{\phi}{2}}. \label{qmb} 
\end{equation}
The target state can be obtained with constant probability by $O(1/P_{\rm m})$ times repetitions of the general algorithm. Hence the total query complexity of the algorithm becomes 
\begin{equation}
Q = \frac{q_{\rm m}}{P_{\rm m}} = \frac{q_{\rm m}}{\beta^{2}} = \frac{\pi}{4\alpha}\frac{B^{3}\sin^{2}\frac{\phi}{2}}{\sin 2\eta}\ .
\end{equation} 
The minimum required number of queries by any quantum algorithm is $O(1/\alpha)$ and hence the general algorithm becomes inferior to the optimal algorithm if and only if $\sin 2\eta \ll 1$ which is true if $\cot 2\eta \gg 1$. By definition, this is true when $A \gg 2\alpha B$. Thus $A = O(\alpha B)$ is a necessary condition for the success of algorithm and as typically $\alpha \ll 1$, $A$ must be close to zero. This is a very restrictive condition. In next section, we introduce a modification of the general algorithm which helps in getting rid off this condition.

\section{MODIFIED GENERAL ALGORITHM}

To get the basic idea behind modification, we note that $O(1/P_{\rm m})$ times repititions of the general algorithm to boost the success probability to a constant value is basically a classical and inefficient process. In quantum setting, a far efficient method is available in the form of quantum amplitude amplification (hereafter referred to as QAA). In QAA, the $|u\rangle$ state is evolved to the target state $|t\rangle$ by $O(1/\beta)$ iterations of the QAA operator $I_{t}I_{u}$ on $|u\rangle$. By definition of the $|u\rangle$ state, we have
\begin{equation}
I_{u} = \mathcal{S}^{q_{\rm m}}I_{s} \mathcal{S}^{-q_{\rm m}}. \label{uintermofs}
\end{equation}
Thus implementation of $I_{u}$ requires implementation of $I_{s}$. The question is that though $I_{t}$ can be implemented easily, the same is not true for $I_{s}$ and the entire motivation behind the construction of the general algorithm started with the hypothesis that we have only the operator $D_{s}$ available and not $I_{s}$. This is what prevents us to use QAA to get the target state $|t\rangle$. We point out that $I_{s}$ is not easily implementable in cases of physical interest~\cite{spatial,fastersearch,fastergeneral,clause,kato,shenvi,realambainis} 

To understand the modification of the general algorithm, we closely examine the possibility of implementing $I_{s}$. The $|s\rangle$ state is an eigenstate of the $D_{s}$ operator with a known eigenvalue $1$. Thus the phase estimation algorithm~\cite{phase} (hereafter referred to as PEA) can be used to approximate $I_{s}$, the selective phase inversion of the $|s\rangle$ state, using multiple applications of the operator $D_{s}$. 

In a recent paper (see Section III of ~\cite{postprocessing}, we have presented a detailed algorithm for the approximate implementation of the selective phase inversion of the unknown eigenstates. There, the algorithm is presented to implement the operator $I_{\lambda_{\pm}}$ which is the selective phase inversion of the $|\lambda_{\pm}\rangle$ subspace of the operator $\mathcal{S}$. Note that there we have considered only a special case of the operator $\mathcal{S} = D_{s}I_{t}^{\phi}$ when $\phi$ is $\pi$ and $D_{s}$ is such that $\Lambda_{1}$ is zero. We have shown there that due to the assumption $|\lambda_{\pm}| \ll \theta_{\rm min}$, the operator $I_{\lambda_{\pm}}$ can be approximated with an error of $\epsilon$ using $O(\ln \epsilon^{-1}/\theta_{\rm min})$ applications of $\mathcal{S}$. 

This is straightforward to extend the same ideas for approximate implementation of $I_{s}$ as by definition, all eigenstates of $D_{s}$ orthogonal to $|s\rangle$ have eigenphases greater than $\theta_{\rm min}$ and $|s\rangle$ is the only eigenstate satisfying $\theta_{s} = 0 \ll \theta_{\rm min}$. Thus $I_{s}$ can be implemented with an error of $\epsilon$ using $O(\ln \epsilon^{-1}/\theta_{\rm min})$ applications of $D_{s}$.

In the general algorithm, the application of $I_{t}$ as well as $D_{s}$ takes unit time step. Thus the operator $\mathcal{S}$ takes two time steps and Eq. (\ref{uintermofs}) implies that $T[I_{u}]$ is $2q_{\rm m} + T[I_{s}]$ where $T[X]$ denotes the time steps needed to implement the operator $X$. The discussion of the previous paragraph implies that
\begin{equation}
T[I_{t}I_{u}] = 2q_{\rm m} + 1+T[I_{s}] = O\left(q_{\rm m } +\frac{\ln \epsilon^{-1}}{\theta_{\rm min}}\right).
\end{equation}
Here $\epsilon$ is the desired error in implementation of $I_{s}$. For QAA, we need $O(1/\beta)$ applications of the operator $I_{t}I_{u}$ and hence same number of the approximate implementations of $I_{s}$. Thus the desired error in each approximate implementation of $I_{s}$ is $O(\beta)$. The total time complexity of the algorithm is then
\begin{equation}
\frac{1}{\beta}O\left(q_{\rm m} + \frac{\ln \beta^{-1}}{\theta_{\rm min}}\right). \label{totaltime1}
\end{equation} 

Let us assume for a moment that
\begin{equation}
\frac{\ln \beta^{-1}}{\theta_{\rm min}} \not\gg q_{\rm m}. \label{assumption}
\end{equation} 
Then the second term in Eq. (\ref{totaltime1}) can be ignored and the total time complexity of the algorithm becomes 
\begin{equation}
O\left(\frac{q_{\rm m}}{\beta}\right) = \frac{\pi B^{2}}{4\alpha} \sin \frac{\phi}{2},
\end{equation}
where we have used Eq. (\ref{qmb}). Thus as desired, the time complexity is completely independent of $\eta$ and $A$. As typically $B$ and $\phi$ are $\Omega(1)$, the time complexity is close to the optimal performance of $O(1/\alpha)$.

The only condition to be satisfied by the algorithm is the assumption (\ref{assumption}). Ignoring the logarithmic factor and using Eq. (\ref{qmb}), the assumption becomes
\begin{equation}
\theta_{\rm min} \not\ll \frac{1}{q_{\rm m}} = \frac{4\alpha}{\pi B}\frac{1}{\sin 2\eta} \approx \frac{4\alpha}{\pi B}\frac{A}{2\alpha B} = \frac{2A}{\pi B^{2}},
\end{equation}
where we have used Eq. (\ref{solutions}) and the fact that $1/\sin 2\eta \approx \cot 2\eta$ whenever $\sin 2\eta \ll 1$. Note that if $\sin 2\eta \not\ll 1$ then there is no need to modify the general algorithm as the original general algorithm is also fast enough. The above condition can be rewritten as
\begin{equation}
A \not\gg 1.57 B^{2}\theta_{\rm min}.
\end{equation}
We compare it with the condition $A \not\gg 2\alpha B$ required for the success of the original general algorithm. As typically $B$ is $\Theta(1)$ and $\theta_{\rm min} \gg \alpha$, the condition for the modified general algorithm is significantly relaxed compared to that for the original general algorithm. 

\section{DISCUSSION AND CONCLUSION}

We have shown that a modification of the general algorithm allows us to get a successful quantum search algorithm using a more general class of diffusion operators. The modification crucially depends upon the phase estimation algorithm and hence the quantum fourier transform. 

A very important application of this modification is in tackling errors in diffusion operators. The original condition $A \not\gg 2\alpha B$ is a very restrictive condition as typically $\alpha$ is a very small quantity. Hence even minor deviations in the diffusion operator can cause failure of the original general algorithm. But the modified general algorithm is robust to such kind of small errors as this allows $A$ to be as large as $O(\theta_{\rm min}B^{2})$. This is a big relief as for typical diffusion operators, the quantity $\theta_{\rm min}$ is much bigger than $\alpha$.

We hope that this modification will help us in designing fast quantum search algorithms under more general situations.

\end{document}